\documentclass[12pt]{article}
\usepackage[numbers]{natbib}
\usepackage{graphicx,color}
\usepackage{times}
\usepackage{geometry}
\usepackage{placeins}
\usepackage[utf8]{inputenc}
\usepackage{enumitem,amssymb}
\usepackage{ragged2e}
\newlist{thematic}{itemize}{8}
\setlist[thematic]{label=$\square$}
\usepackage{pifont}
\newcommand{\cmark}{\ding{51}}%
\newcommand{\done}{\rlap{$\square$}{\raisebox{2pt}{\large\hspace{1pt}\cmark}}%
\hspace{-2.5pt}}

\newcommand {\Chandra}{{\it Chandra }}
\newcommand {\XMM}{{\it XMM-Newton }}

\definecolor{rubinered}{rgb}{0.82, 0.0, 0.34}

\usepackage{hyperref}

\begin{document}
\raggedright
\huge
Astro2020 Science White Paper \linebreak

\centering{
Probing 3D Density and Velocity Fields}\\
\centering{of ISM in Centers of Galaxies}\\
\centering{with Future X-Ray Observations}\\
%\linebreak
\normalsize

\noindent \textbf{Thematic Areas:} \hspace*{60pt} $\square$ Planetary Systems \hspace*{10pt} \done  Star and Planet Formation \hspace*{20pt}\linebreak
$\square$ Formation and Evolution of Compact Objects \hspace*{31pt} $\square$ Cosmology and Fundamental Physics \linebreak
  $\square$  Stars and Stellar Evolution \hspace*{1pt} $\square$ Resolved Stellar Populations and their Environments \hspace*{40pt} \linebreak
  \done \, Galaxy Evolution   \hspace*{45pt} $\square$             Multi-Messenger Astronomy and Astrophysics \hspace*{65pt} \linebreak
  
\textbf{Authors:}

Name:	E.Churazov,$^{1,2}$ I.Khabibullin,$^{1,2}$ R.Sunyaev,$^{1,2}$ A.Vikhlinin,$^{3}$ G.Ponti,$^{4}$ C.Federrath,$^{5}$ S.Walch$^{6}$
 \linebreak	
 
$^1$ Max-Planck-Institut f\"ur Astrophysik, Karl-Schwarzschild-Strasse 1, 85741
Garching, Germany\\
$^2$ Space Research Institute (IKI), Profsoyuznaya 84/32, Moscow 117997, 
Russia\\
$^3$ Harvard-Smithsonian Center for Astrophysics, 60 Garden St.,
Cambridge, MA 02138, USA \\ 
$^4$ INAF – Osservatorio Astronomico di Brera, via E. Bianchi 46, I-23807, Merate, Italy\\
$^5$ Research School of Astronomy and Astrophysics, Australian National University, Canberra, ACT 2611, Australia\\
$^6$ I. Physikalisches Institut, Universit\"at zu K\"oln, Zülpicher Str. 77, D-50937 K\"oln, Germany

\vspace{0.5cm}
%\textbf{Co-authors:}Gabriele Ponti (INAF – Osservatorio Astronomico di Brera, via E. Bianchi 46, I-23807, Merate, Italy)
% \linebreak

\justifying \textbf{Abstract:}
Observations of bright and variable ``reflected" X-ray emission from molecular clouds located within inner hundred parsec of our Galaxy have demonstrated that the central supermassive black hole, Sgr A*, experienced short and powerful flares in the past few hundred years. These flares offer a truly unique opportunity to determine 3D location of the illuminated clouds (with $\sim10$ pc accuracy) and to reveal their internal structure (down to 0.1 pc scales). Short duration of the flare(s), combined with X-rays high penetration power and insensitivity of the reflection signal to thermo- and chemo-dynamical state of the gas, ensures that the provided diagnostics of the density and velocity fields is unbiased and almost free of the projection and opacity effects. Sharp and sensitive snapshots of molecular gas accessible with aid of future X-ray observatories featuring large collecting area and high angular (arcsec-level) and spectral (eV-level) resolution cryogenic bolometers will present invaluable information on properties of the supersonic turbulence inside the illuminated clouds, map their shear velocity field and allow cross-matching between X-ray data and velocity-resolved emission of various molecular species provided by ALMA and other ground-based facilities. This will highlight large and small-scale dynamics of the dense gas and help uncovering specifics of the ISM lifecycle and high-mass star formation under very extreme conditions of galactic centers. While the former is of particular importance for the SMBH feeding and triggering AGN feedback, the latter might be an excellent test case for star formation taking place in high-redshift galaxies.

%which are not uncommon to the conditions characteristic to star-forming galaxies at high redshift.

%Currently available data  allow us to estimate age and fluence of the most recent flare, and to construct crude 3D gas density maps in a 3-4~pc thick slice.
%Future sensitive X-ray observatories featuring cryogenic bolometers with high angular resolution will be able to reconstruct the gas density probability distribution function (a proxy to basic properties of supersonic turbulence), map the shear velocity field, and cross-match X-ray data with the velocity-resolved emission of various molecular species. 
%This will be of crucial importance for revealing large and small scale dynamics of the dense gas, shedding light on specifics of the ISM lifecycle and high mass star formation under very extreme conditions of the Galactic Center region.

\thispagestyle{empty}
\setcounter{page}{0}

\pagebreak

%Insert your white paper text here (max of five pages including figures).

\section{Molecular Clouds in the Galactic Center}

~~~~The cold phase of the interstellar medium condenses into dense and compact clouds, containing predominantly self-gravitating molecular gas shaped by supersonic turbulence \citep{2001RvMP...73.1031F,2004RvMP...76..125M,McKee2007,2014prpl.conf...77P}. Fragmentation of this gas into collapsing cores seeds pre-stellar objects, leading to star formation and its feedback on the parent environment \citep{2015MNRAS.450.4035F,2018PhT....71f..38F}. Thus, molecular clouds compose a key part of the ISM life-cycle since they facilitate an extremely important link between global ($\gtrsim100$ pc) gas flows and cradles of individual massive stars ($\lesssim0.01$ pc) \citep{2015MNRAS.454..238W}.

Molecular clouds in the Central Molecular Zone (CMZ) of our Galaxy are of particular interest in this regard, since they evolve in a very extreme environment of the Galactic Center, and also because their measured star formation efficiency appears to be an order of magnitude lower compared to the molecular clouds in the Galactic disk \citep{2017A&A...603A..89K}.  Additionally, these clouds trace the flows of the cold gas that might be responsible for feeding of the Galaxy's central supermassive black hole, Sgr A*, and hence regulating its feedback.  

The internal structure of molecular clouds is determined by a complex, scale-dependent interplay between several processes, which are far from being fully understood despite their fundamental importance for a range of astrophysical contexts \citep[][]{McKee2007,Heyer2015}. Namely, driving and decay of the supersonic turbulence, generation of magnetic fields and cosmic ray propagation are all likely to play a role in the mass and energy flows across a range of scales spanning many orders of magnitude. On top of this, molecular clouds are essentially dynamical structures evolving as their internal life-cycle proceeds as well as in response to changes in external conditions \citep[e.g.][]{2015MNRAS.447.1059K}. In that sense, molecular clouds in the CMZ are subject to the most extreme and dynamic environment in the Milky Way, so they are ideal laboratories to study the aforementioned physical processes in action \citep[][]{2017A&A...603A..89K}.     

For a typical molecular cloud, quasi-isothermal supersonic turbulence is believed to shape the density field at $1-10$ pc scales, resulting in the log-normal distribution function \citep{1994ApJ...423..681V} with the width determined by the Mach number of the turbulent motions and the nature of forcing \citep{2008ApJ...688L..79F}. On sub-pc scales, self-gravity of individual dense cores starts to take over, marking the transition to the coherent (rather than turbulent) motions \citep[e.g.][]{Goodman1998}, and it is these scales that are believed to be crucial for determining a cloud's overall massive star-formation efficiency \citep[e.g.][]{Williams2000,Ward2007,2018MNRAS.479.1702O,2019arXiv190200934K}. Thus, probing a range of scales from 10 pc down to 0.1 pc is absolutely necessary to reconstruct the physical picture of the multi-scale gas dynamics by separating various effects and revealing their interconnections.

Physical conditions and hence emission properties of the gas vary noticeably from scale to scale, so it is very hard to construct a probe that would be unbiased and free of opacity and projection effects. In contrast to the nearby molecular complexes, any emission from the clouds in the CMZ is strongly affected by interstellar absorption and by the fact that the typical angular resolution of a single-dish antenna ($\sim$ 30 arcsec) allows resolving only scales above 1.5 pc at the distance to the Galactic Center. Single-dish observations however provide information about the total mass of the cloud (hence its average density) and velocity dispersion (hence the Mach number of the gas motions inside it). Interferometric observations (e.g. with ALMA) are capable of resolving structures down to $ \sim 1$ arcsec (and even smaller) resolution, so that individual filaments and dense cores can be efficiently detected \citep{2015ApJ...802..125R,2017IAUS..322..162U}. Reconstruction of the density PDF with such data is however problematic, due to leakage of the large-scale power \citep{2016ApJ...832..143F}.

Fortunately, in the CMZ the Nature offers us a unique diagnostic tool - illumination of the clouds with the Sgr~A* flare, that is short enough to remove all adverse projection or opacity effects in the "reflected" signal. This X-ray reflection technique is very much complementary to both single-dish and interferometric observations in molecular line emission, basically providing a link between them.

%%%%%%%%%%%%%%%%%%%%%%%%%%%%%%%%%%%%%%%%%%%%%%%%%%%%%%%%%%%%%%%%%
\begin{figure}
\includegraphics[trim= 5cm 2cm 5cm 1cm,width=0.28\textwidth]{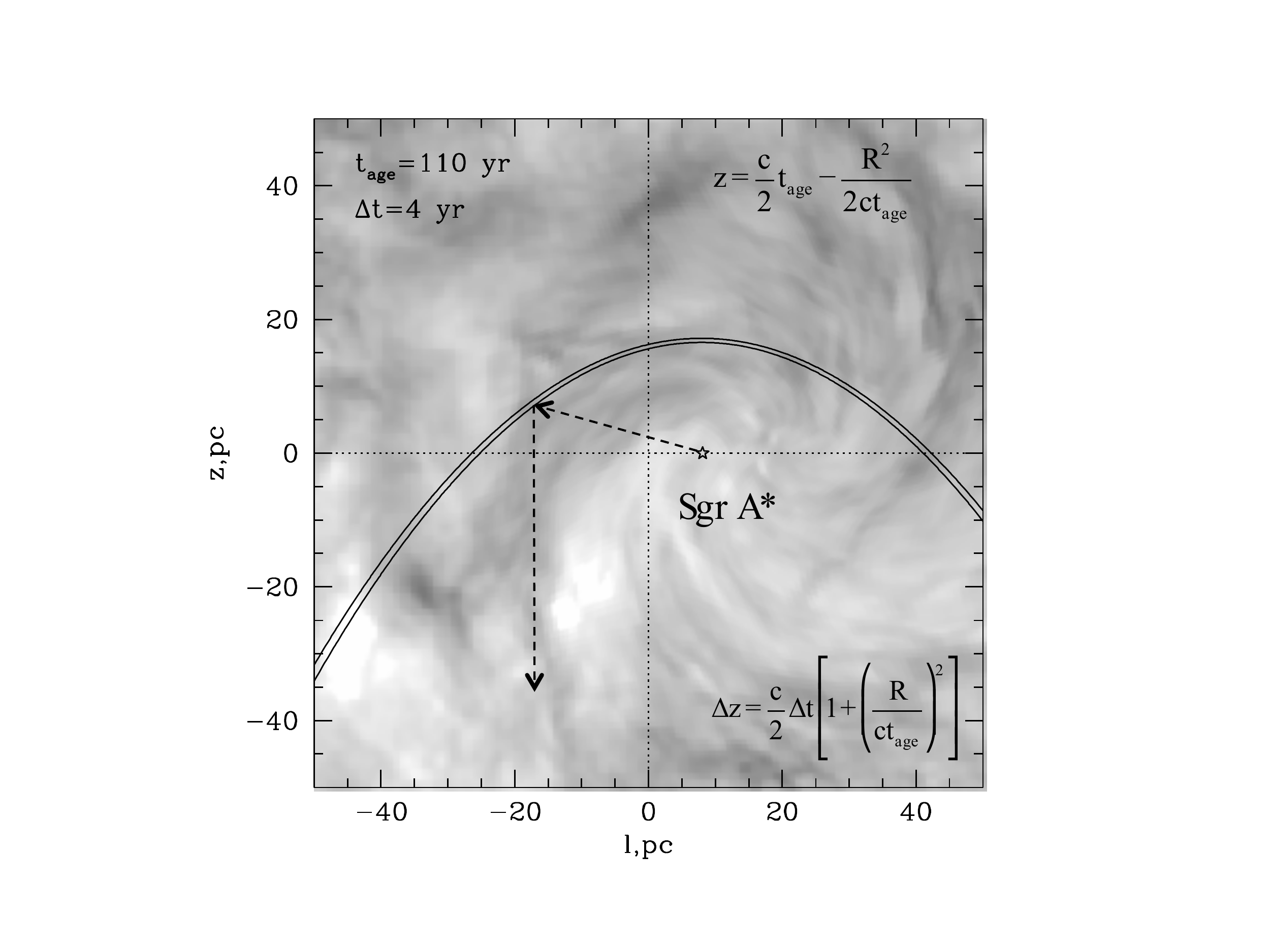}
\includegraphics[trim= 0cm -2.2cm 0.cm 0cm, width=0.8\textwidth,clip=t,angle=0.,scale=0.46]{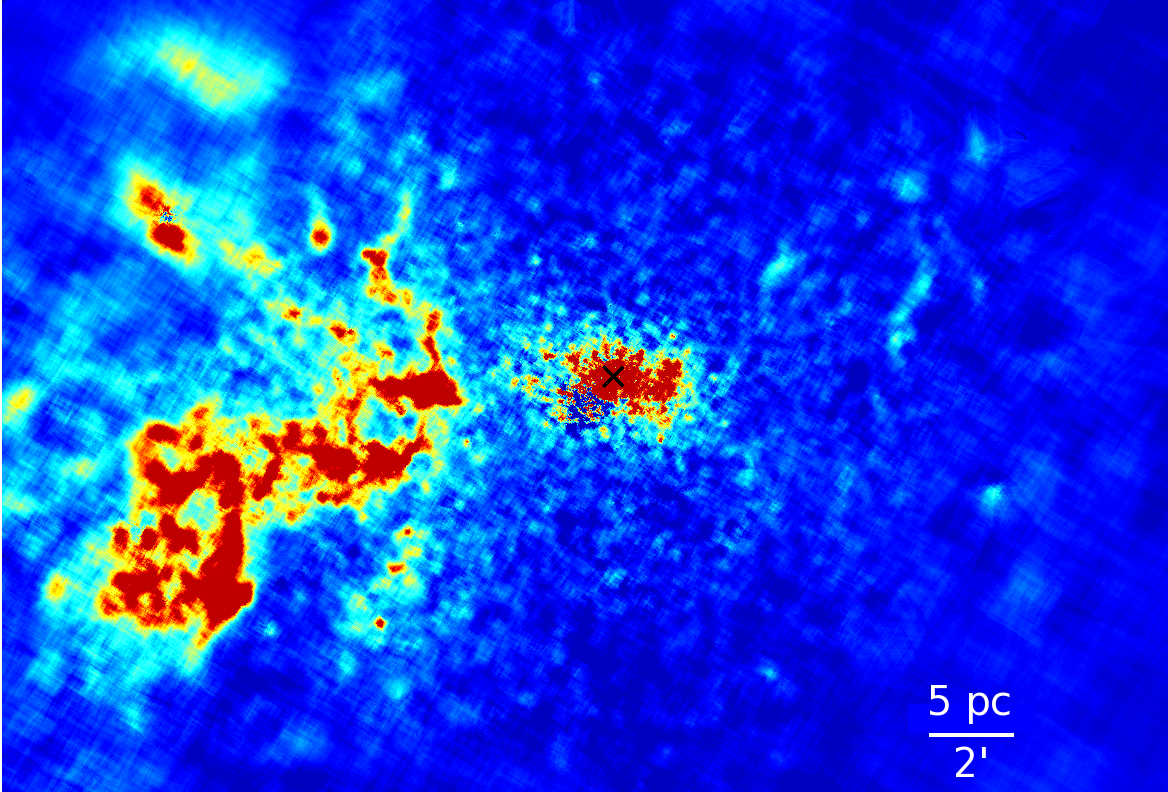}
\includegraphics[trim= 0mm -1.4cm 0.5cm 5cm, width=1\textwidth,clip=t,angle=0.,scale=0.3]{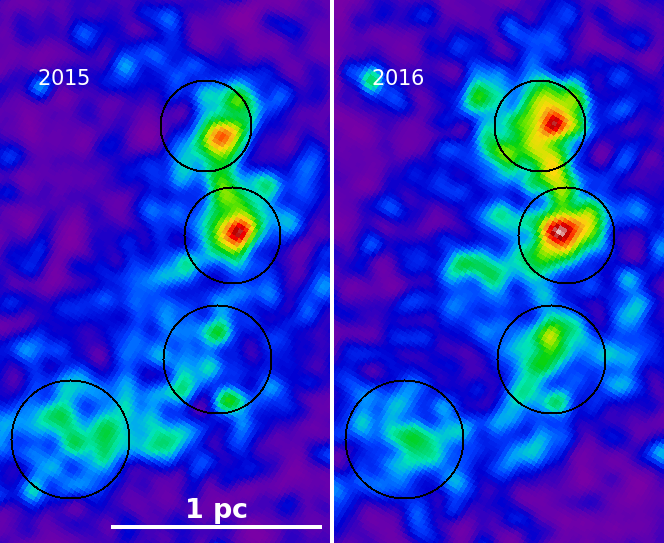}
\caption{\small {\bf Left:} Sketch of molecular gas in the Galactic Center region exposed to a 4 years long flare
  that happened 110 years ago (a view from above the Galactic Plane). The light propagates from Sgr~A* to the dense gas and then to the observer \citep{1998MNRAS.297.1279S}.  The locus of illuminated gas is a space between the two ellipsoids. Only this thin slice should be visible in X-rays, eliminating all adverse effects of projection. {\bf Middle:} \Chandra image of the reflected component in the central region of the Galaxy. This component can be easily separated from other background and foreground components using its distinct spectral shape.
  \textbf{Right}: Changes in the 4-8 keV flux maps on a time scale of one year in a small patch $\sim20$~pc from Sgr~A*. 
  %{\bf Right:} Reconstructed 3D distribution of the molecular gas density from the \XMM data accumulated over 15 years of observations. The width of the reconstructed layer is $\sim 3.5\;{\rm pc}$ and the radius is $\sim 35\;{\rm pc}$; adapted from \citep{2017MNRAS.465...45C}). Future X-ray observatories should be able to probe the PDF of the gas density in the molecular gas on small scales and the molecular gas velocity field.
\label{fig:sketch}
}
\end{figure}
%%%%%%%%%%%%%%%%%%%%%%%%%%%%%%%%%%%%%%%%%%%%%%%%%%%%%%%%%%%%%%%%%

\section{Illumination of molecular clouds by Sgr~A* X-ray flare}

~~~~The supermassive black hole Sgr A* at the center of the Milky
Way is currently very dim \citep{2010RvMP...82.3121G}, although it has
experienced powerful outbursts  in the recent past,
as revealed by reflected X-ray emission coming from dense
molecular clouds with a light-travel-time delay of tens to hundreds of years \citep{1993ApJ...407..606S,1996PASJ...48..249K,2004A&A...425L..49R,2010ApJ...719..143T,2015ApJ...814...94M,2017MNRAS.468.2822K}. One recent outburst took place
$t_{age}\sim 110~{\rm yrs}$ ago, when Sgr A* was more than a million times brighter than today for a period of time 
shorter than $\Delta t\sim$ 1.6 years \citep{2017MNRAS.465...45C,2017MNRAS.468..165C,2017MNRAS.471.3293C}.
The upper limit on the flare duration is set by variability of the observed emission \citep[e.g.][]{2007ApJ...656L..69M,2010ApJ...714..732P,2013A&A...558A..32C}. The total energy emitted during such a flare amounts to  $\sim$ few $ 10^{47}$~erg. 

The nature of such flare remains unclear (it might be a signature of tidal disruption of a planet followed by a short episode of debris accretion onto the supermassive black hole). Clues could be \textbf{found} by recovering the {\bf light curve of the flare}, which is possible if a very compact knot is \textbf{found} in the illuminated molecular gas. If the light-crossing time of the knot is shorter than the duration of the flare, then the variation of its reflected emission accurately reproduces the flare profile. A comparison of light curves for several compact knots at different locations would be useful to verify if there was a single flare or many flares separated by extended quiescent periods.     

\begin{figure}
\includegraphics[trim= 0cm 6cm 0.cm 4cm, width=0.65\textwidth,clip=t,angle=0.,scale=0.5]{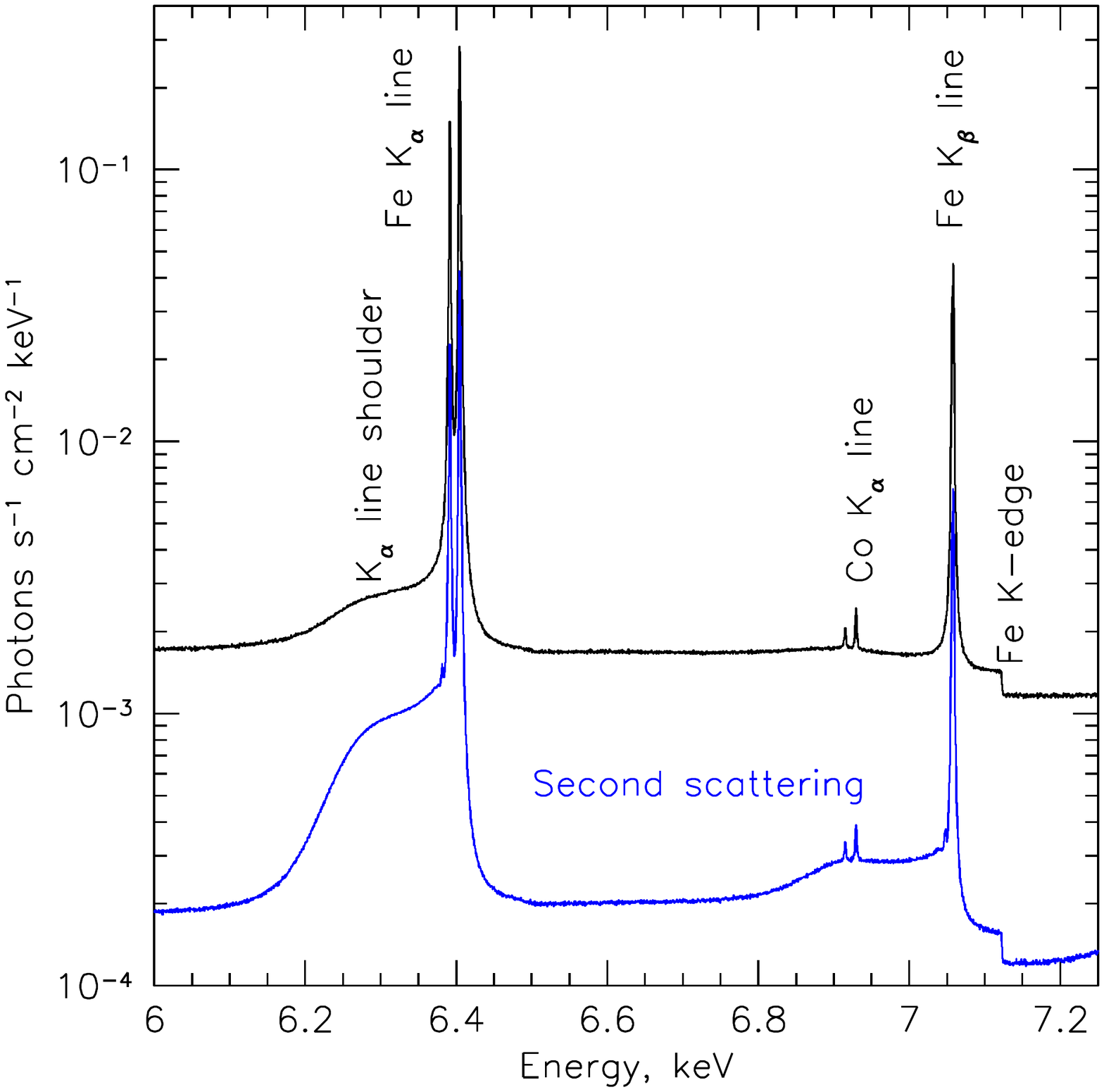}
\includegraphics[trim= 0cm 6cm 0.cm 4cm, width=0.65\textwidth,clip=t,angle=0.,scale=0.5]{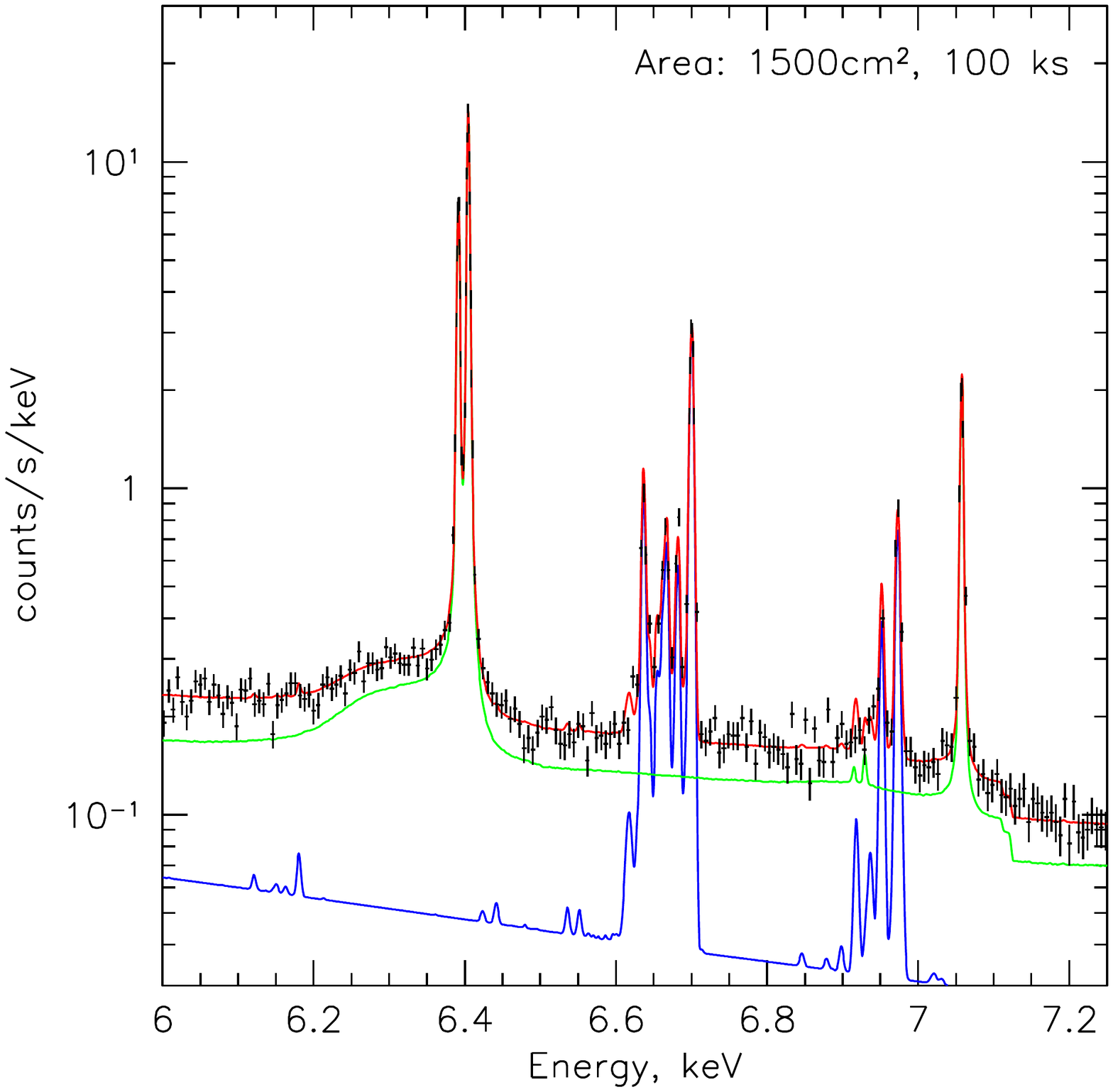}
\includegraphics[trim= 0mm 6cm 0.cm 4cm, width=0.65\textwidth,clip=t,angle=0.,scale=0.5]{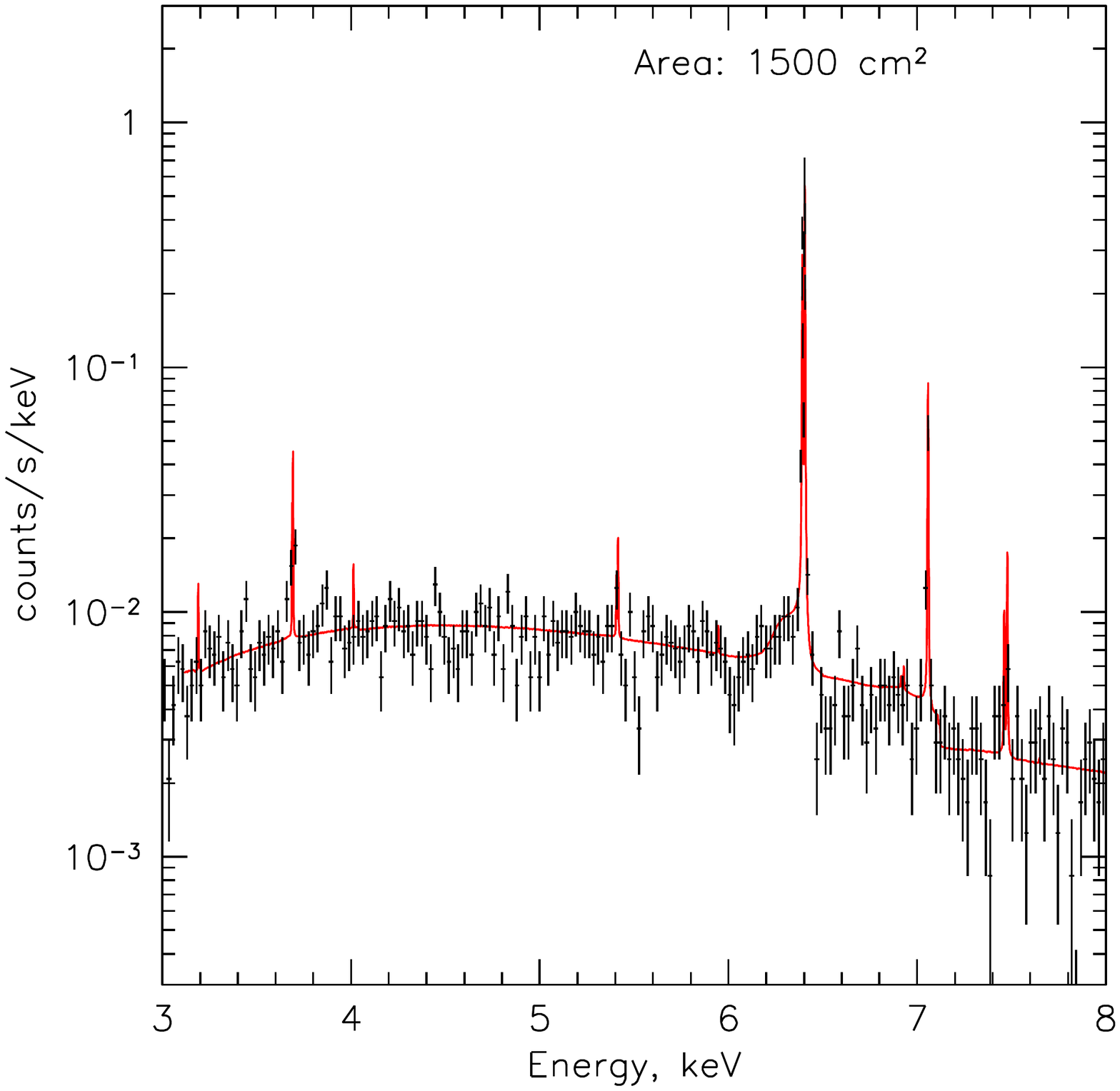}
\caption{\small  {\bf Left:} Predicted spectrum arising from a slab of cold gas with the Thomson optical depth $\tau_T=0.2$ illuminated by a power-law X-ray continuum. Only a narrow region near the prominent iron 6.4 keV line is shown (black line). The blue line shows a doubly-scattered emission profile. This emission can be observed even after the primary photons have already left the cloud. {\bf Center:} Predicted reflected emission, which will be observed with future X-ray observatories having $\sim 1500\;{\rm cm^2}$ effective area and $\sim 3$~eV energy resolution from the bright region shown in Fig.~1 (middle panel) in 100~ks observation. The blue curve corresponds to a thermal plasma emission, schematically illustrating the contribution of unresolved compact sources. \textbf{Right}: The simulated spectrum from one of the knots shown in the right panel of Fig.~\ref{fig:sketch} for an instrument with 10 times larger effective area than \Chandra and an energy resolution of 3~eV. 
\label{fig:shoulder}
}
\end{figure}

\begin{figure}
\href{https://wwwmpa.mpa-garching.mpg.de/~churazov/gc_rot_1.mp4}{\includegraphics[trim= 1cm 0.cm 1.cm 0cm, width=0.7\textwidth,clip=t,angle=0.,scale=0.4]{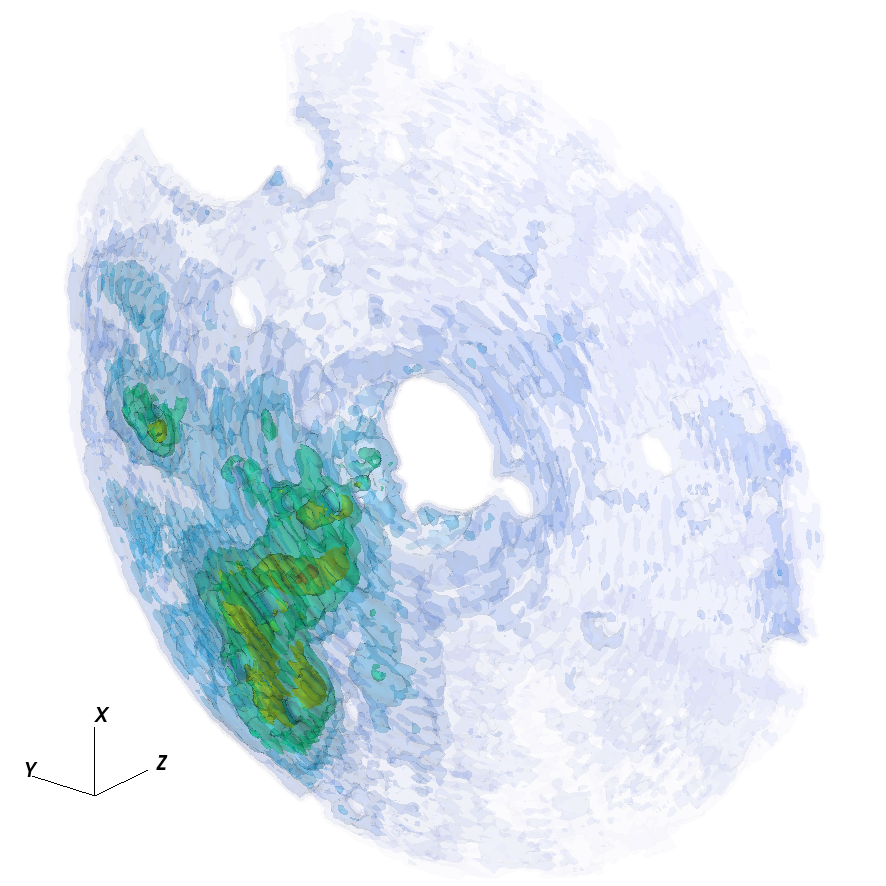}}
\includegraphics[trim= 0cm 7cm 0.cm 5cm, width=0.72\textwidth,clip=t,angle=0.,scale=0.5]{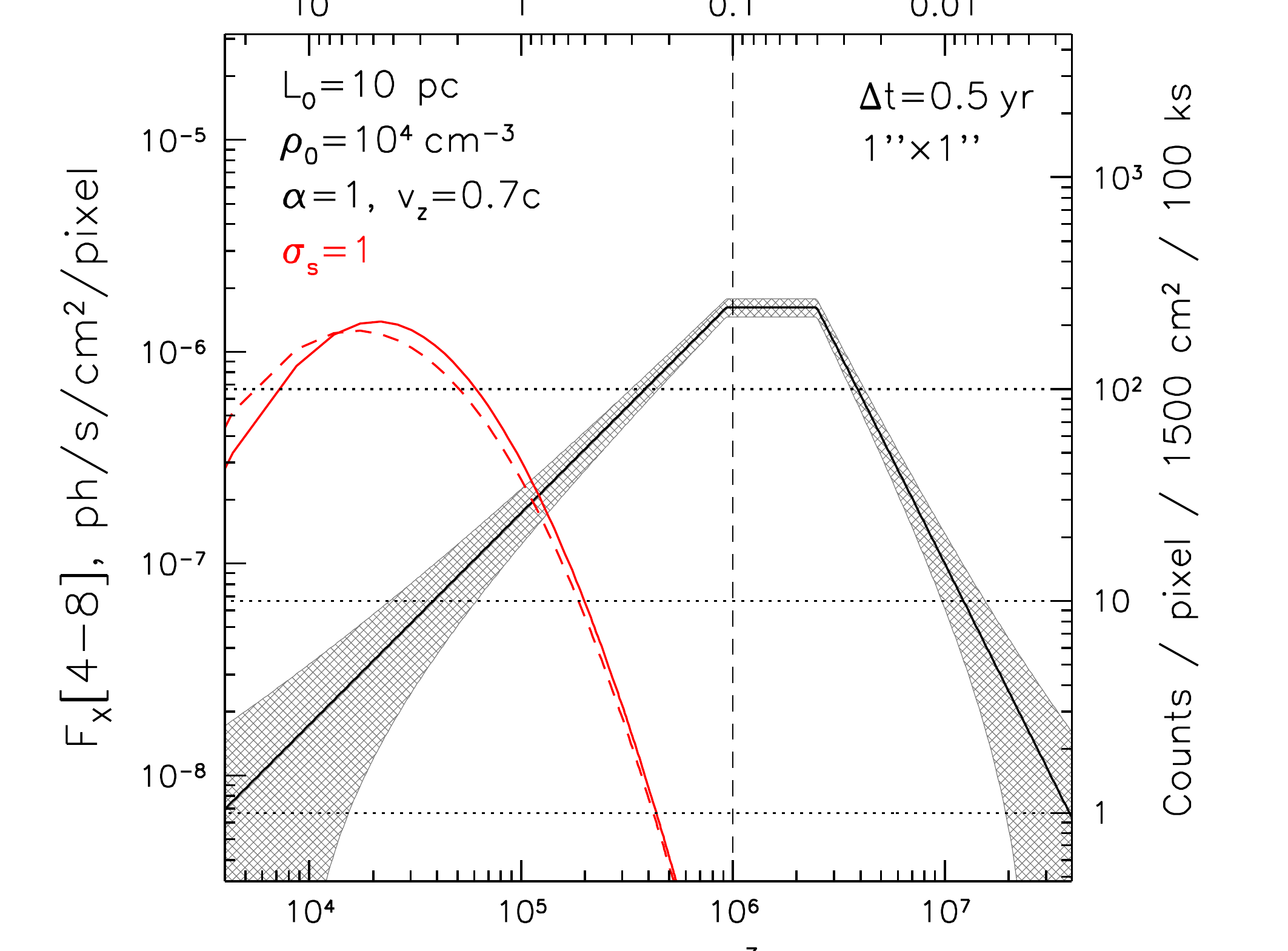}
\includegraphics[trim= 1cm 7cm 0.cm 7cm, width=0.73\textwidth,clip=t,angle=0.,scale=0.5]{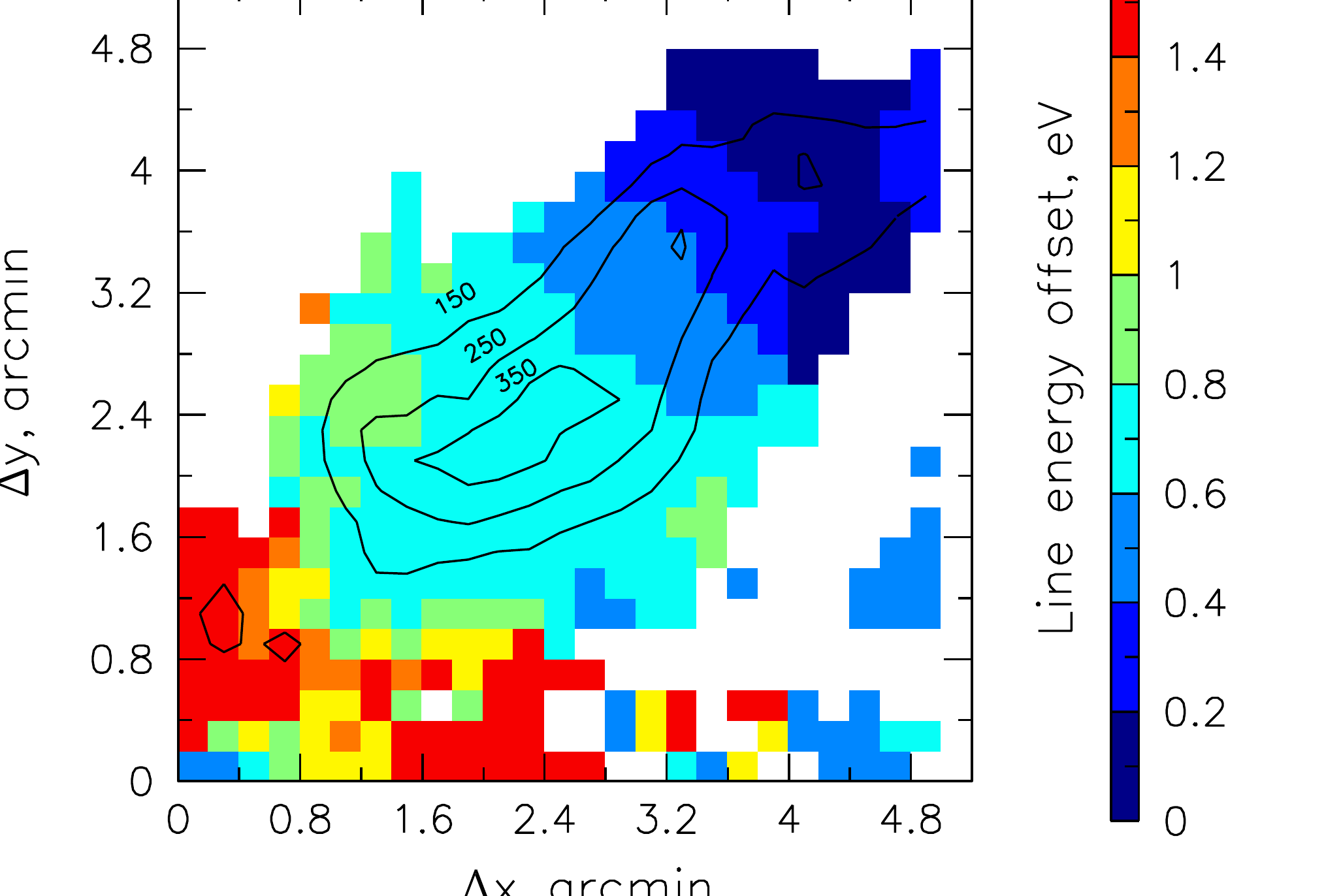}
\caption{\small  {
{\bf Left:} Reconstructed \href{https://wwwmpa.mpa-garching.mpg.de/~churazov/gc_rot_1.mp4}{\color{blue}{3D distribution}} of the molecular gas density from the \XMM data accumulated over 15 years of observations. The width of the reconstructed layer is $\sim 3.5\;{\rm pc}$ and the radius is $\sim 35\;{\rm pc}$; adapted from \citep{2017MNRAS.465...45C}. %Future X-ray observatories should be able to probe the PDF of the gas density in the molecular gas on small scales and the molecular gas velocity field.
\bf Center:} Linearity of the surface brightness of reflected X-ray emission and gas density (and characteristic size) of a substructure inside a molecular cloud with power-law relation between density (bottom axis) and size (top axis) in a form $ \rho\propto l^{-\alpha} $ with $\alpha=1$, and $\rho_0=10^4$ cm$^{-3}$ at $L_0=10$ pc.  Front propagation speed along the line-of-sight is set equal to $v_z=0.7c$, the flare's duration is $ \Delta t=0.5$ yr. The right axis shows the expected number of counts per 1''$\times$1'' pixel. The red dashed curve (arbitrary normalization) shows mass distribution between the scales for a cloud having log-normal density distribution with mean density $\rho_0=10^4$ cm$^{-3}$ and dispersion $\sigma_s=1$, while the solid red line shows how this PDF will be distorted by the Eddington bias corresponding to $\approx2$ counts per pixel at $\rho_0=10^4$ cm$^{-3}$. {\bf Right:} Simulated shift of the line 6.4 keV centroid for a shear velocity field resembling observed average densities and velocity gradients in the Brick cloud based on HNCO data of the MOPRA survey \cite{2015ApJ...802..125R,2016ApJ...832..143F}. 
%The black contours show the number of line counts to be detected after a 100 ks exposure by an instrument with effective area $1500$ cm$^2$ at 6.4 keV.  
\label{fig:pdf}
}
\end{figure}

{\bf The geometry of the illuminated region} is uniquely set by the time delay due to propagation of scattered light from Sgr~A* to the cloud and then to the observer (see Fig.~\ref{fig:sketch}). Hence, knowing the age of the outburst gives an opportunity to reconstruct the {\bf 3D location of the reflecting molecular gas} with respect to Sgr~A* (see Fig.~\ref{fig:sketch} and also the left panel in Fig.~\ref{fig:pdf}). Moreover, knowing that the outburst was shorter than a few years allows one to turn X-ray reflection into an extremely powerful tool for probing characteristics of the illuminated molecular gas itself. Indeed, the line-of-sight thickness of the illuminated region is just $\Delta z \sim c\Delta t\sim0.2$~pc (and possibly even narrower), i.e. much smaller then the typical size of the whole cloud ($\sim$10 pc). Thus, the outburst effectively `cuts out' a {\bf thin 2D slice of molecular gas} at any given moment, and surface brightness of the X-ray emission is directly proportional to the local gas density (see Figs.~\ref{fig:sketch} and~\ref{fig:pdf}). The reflected X-ray emission has a very characteristic spectral shape (Fig.~\ref{fig:shoulder}), which facilitates its clean separation from other background and foreground components. As a result, the observed X-ray surface brightness is linked to the underlying density field of the molecular gas in a straightforward fashion, so it can readily be exploited to derive statistical properties of the latter.

The first attempt of such a study \citep{2017MNRAS.465...45C} used \Chandra data to reconstruct the {\bf gas density PDF}. Although the measured distribution function was consistent with the expectations for quasi-isothermal supersonic turbulence, i.e. having log-normal shape with the width possibly indicating mostly solenoidal driving \citep{2017MNRAS.465...45C}, it has been recognized that the PDF measured this way suffers from substantial statistical issues both at the low-surface brightness end {(Eddington bias)} and for the brightest individual pixels (shot noise and sampling variance). These problems can be cured with increased exposure \citep{2019K}. However, simply accumulating the data taken at different epochs causes another problem -- all sub-pc structures are expected to be highly variable (see Fig.\ref{fig:sketch}) and co-adding the images smears out the smallest scales. In principle, the \Chandra angular resolution ($1''$ corresponds $\sim$0.04~pc) potentially offers the \textbf{dynamic range of some 200} in resolved spatial scales. Useful constraints can still be achieved with a Ms-long exposure, (collected over the period shorter than $\sim$4 months), but the \Chandra effective area is simply too small to exploit this diagnostic fully, while the angular resolution of \XMM limits the dynamic range. A natural solution - a telescope with the effective area ten times larger than  \Chandra and a comparable angular resolution.

%%%%%%%%%%%%%%%%%%%%%%%%%%%%%%%%%%%%%%%%%%%%%%%%%%%%%%%%%%%%%%%%%
%%%%%%%%%%%%%%%%%%%%%%%%%%%%%%%%%%%%%%%%%%%%%%%%%%%%%%%%%%%%%%%%%

A further nontrivial diagnostic can be provided by measurements of the total flux and time variations of the {\bf fluorescent line Compton shoulder}. For a compact cloud entirely illuminated by the flare, the spectrum should be similar to the one shown in Fig.~\ref{fig:shoulder}, where the flux in the line shoulder $F_{s}$ is approximately $\tau_{c}$ times smaller than the flux in the narrow fluorescent lines $F_{l}$. However, for a larger cloud, whose light-crossing time is longer than the duration of the flare, the  strength of the shoulder should grow steadily with time as the flare front propagates through the cloud, reaching maximum values of $\tau_{c}\times F_{l}$ by the time when the front is about to leave the cloud. Once the front leaves the cloud, the ratio $\frac{F_{s}}{F_{line}}$  jumps to a level of order unity - an unambiguous spectral signature of the second (and higher-order) scattering \citep{1998MNRAS.297.1279S,2016A&A...589A..88M}. The life-time of the second scattering component is set by the light crossing time of the cloud or of the scattering environment on larger scales and, therefore, probes the mean gas density of the entire region and, possibly, low-density extended envelopes of molecular clouds connecting them with the ambient ISM.  

%%%%%%%%%%%%%%%%%%%%%%%%%%%%%%%%%%%%%%%%%%%%%%%%%%%%%%%%%%%%%%%%%
%\begin{figure*}
%\begin{center}
%\includegraphics[trim= 0mm -1.2cm 0.5cm 5cm, width=1\textwidth,clip=t,angle=0.,scale=0.33]{bright_filament_y15_y16.png}
%\includegraphics[bb=20 165 580 705, width=1\textwidth,clip=t,angle=0.,scale=0.31]{spec_s1.pdf}
%\includegraphics[trim= 0mm 4.5cm 0.5cm 5cm, width=1\textwidth,clip=t,angle=0.,scale=0.31]{spec_s1.pdf}
%\includegraphics[trim= 0mm 4.5cm 0.5cm 5cm, width=1\textwidth,clip=t,angle=0.,scale=0.31]{spec_blob.pdf}
%\caption{\small
%\textbf{Left}: Changes in the 4-8 keV flux maps on a time scale of one year in a small patch $\sim20$~pc from Sgr~A*.  
%\textbf{Center}: The 2016 \Chandra spectrum of one of the knots (top region in the left panel). The total number of counts is $\sim$300 in the 4-8 keV band. The black solid line is the reflection model spectrum. \textbf{Right}: The simulated spectrum from one of the knots shown in the right panel of Figure \ref{} for an instrument with 10 times larger effective area than \Chandra and an energy resolution of 3~eV. 
%\label{fig:filament}
%}
%\end{center}
%\end{figure*}
%%%%%%%%%%%%%%%%%%%%%%%%%%%%%%%%%%%%%%%%%%%%%%%%%%%%%%%%%%%%%%%%%

%%%%%%%%%%%%%%%%%%%%%%%%%%%%%%%%%%%%%%%%%%%%%%%%%%%%%%%%%%%%%%%%%
%%%%%%%%%%%%%%%%%%%%%%%%%%%%%%%%%%%%%%%%%%%%%%%%%%%%%%%%%%%%%%%%%

%\FloatBarrier

The next frontier is the {\bf mapping of the gas velocity field} using the centroid shift of the 6.4 keV line.
Typical turbulent velocities of the order of few ${\rm km\;s^{-1}}$ can hardly be measured even with the next generation of X-ray telescopes. However, the shear velocity in the Galactic Center region is much larger and can be mapped with cryogenic bolometers. This is true not only for clouds separated by tens of pc, but also for the velocity variations within individual clouds (see Fig.~\ref{fig:pdf}).

Having in hand the 3D positions and the line-of-sight velocities, the next natural step is to {\bf cross-match the X-ray data and velocity-resolved emission of various molecular species}. This would allow converting PPV molecular data into a much more powerful dataset, since the line-of-sight positions of clouds will be known. The main limitation of this approach is that only molecular clouds that are illuminated by the flare during the time span of observations can be identified. 

We reiterate here that the reflected X-ray signal above 4~keV is essentially insensitive to variations of temperature of the ionization state of the gas (at the levels characteristic for typical warm or cold phases of the ISM). There are subtle and interesting effects associated with electrons bound in different species, e.g., helium \citep{1996AstL...22..648S}, but such effects are beyond the scope of this paper.

\section{Summary}

~~~~~~Presence of large quantities of molecular gas in the vicinity of Sgr~A* opens the possibility for synergetic studies of the supermassive black hole past outbursts reflected by the gas in clouds and the fundamental properties of the molecular clouds themselves, which are not easily accessible without illumination by these outbursts.  Table~\ref{tab:sum} summarizes the properties that can be constrained and the required capabilities of the telescopes. The full potential of this diagnostic can only be unraveled with the next generation of high-sensitivity, high-energy and high-angular resolution X-ray observatories (see Table~\ref{tab:req}).

Finally, high sensitivity and spectral resolution would allow for a systematic study of the {\bf reflected/delayed emission in other galaxies} \citep{1998MNRAS.301L...1G,2017IAUS..322..253C,2019ApJ...870...69F}, effectively probing the outburst rates (in particular, TDEs) over the last few hundred years in a large sample of galaxies.

\begin{table*}[b!]
\begin{center}
  \caption{\small Road-map for probing Sgr~A* flare and molecular clouds with X-ray data. Driving capabilities are coded as follows: {\bf A} - large effective Area; {\bf I} - excellent Imaging; {\bf S} - excellent Spectral resolution; {\bf T} - possibility to make observations, separated in Time by few months; {\bf P} - Polarimetry in X-ray band. "Now" stands for current generation of X-ray telescopes.} \vspace{0.5cm}
    \label{tab:sum}
  \begin{tabular}{l | l | l}
    \hline
    \textbf{Parameter}     & \textbf{Methods} & \textbf{Driving capabilities} \\
    \hline 
    & {\it Sgr~A* flare diagnostic} & \\
    \hline
    $t_{age}$     & Structure functions in time and space  & Now, {\bf T, I, A, S}\\   
    & Equivalent width of the 6.4 keV line  & Now, {\bf S, A, I} \\
    (also source position) & Polarization  & {\bf P, A, S, I}\\
        \hline
        $\Delta t$     &  Structure functions in time and space & Now, {\bf T, I, A, S}\\   
                  & Smallest clouds (short light crossing time) & {\bf I, A, S}\\   
          \hline
        $L_{X,min}$     & Brightest clouds (maximal density) & {\bf I, A}\\   
          \hline
          $L_X(t)$     & Smallest clouds  (short light crossing time)& {\bf I, A, S}\\             \hline
          $L_{X}\Delta t$   & X-ray images + molecular data & {\bf I, A} \\   
          \hline
              & {\it Molecular gas diagnostic} & \\

        \hline
          $PDF(\rho)$     & Deep X-ray images & {\bf I, A, S}\\
          \hline
          $\rho_{3D}(l,b,z)$   & Multiple X-ray images  & Now, {\bf I, A, S} \\
          \hline
          $v_z(l,b,z)$     & High resolution X-ray spectroscopy  & {\bf S, A, I}\\
                    \hline
                    PPV $\rightarrow$3D      & High resolution X-ray spectroscopy  & {\bf S, A, I}\\
                    \hline
                    Other galaxies     & Sample of quiescent galaxies  & {\bf I, A, S}\\

    \hline
\end{tabular}
\end{center}
\end{table*}

\begin{table*}[b!]
\begin{center}
  \caption{Main requirements for future X-ray observatories} \vspace{0.5cm}
    \label{tab:req}
  \begin{tabular}{l | l}
    \hline
    \textbf{Goal}     & \textbf{Requirements} \\
    \hline
    Resolve 0.05~pc @ GC distance & PSF$\sim$1"  \\
    \hline
    Get $>$200 counts in 6.4 keV line at scales $\sim$0.1~pc (Fig.\ref{fig:pdf}) & Area$\sim 1500\;{\rm cm^2}$ @  6.4 keV\\
    \hline
    Error in velocity $< 10\;{\rm km\;s^{-1}}$   & FWHM $\sim$3 eV  \\
    \hline
\end{tabular}
\end{center}
\end{table*}

\newpage
%\section{References}
%{\footnotesize}
\pagebreak

%\pagebreak
%\textbf{References}

\begin{thebibliography}{99}


\bibitem[\protect\citeauthoryear{Churazov et al.}{2017a}]{2017MNRAS.465...45C} Churazov E., Khabibullin I. et al., 2017, MNRAS, 465, 45 


\bibitem[\protect\citeauthoryear{Churazov et al.}{2017b}]{2017MNRAS.468..165C} Churazov E., Khabibullin I. et al., 2017, MNRAS, 468, 165 


\bibitem[\protect\citeauthoryear{Churazov et al.}{2017c}]{2017MNRAS.471.3293C} Churazov E., Khabibullin I. et al., 2017, MNRAS, 471, 3293 


%\bibitem[Clark et al.(2012)]{Clark2012} Clark, P.~C., et al.\ 2012, MNRAS, 424, 2599 

\bibitem[Clavel et al.(2013)]{2013A&A...558A..32C} Clavel, M., et al.\ 2013, A\&A, 558, A32 

\bibitem[Clavel et al.(2017)]{2017IAUS..322..253C} Clavel M., et al., 2017, IAUS, 322, 253 


%\bibitem[Couderc(1939)]{1939AnAp....2..271C} Couderc, P.\ 1939, AnAp, 2, 
%271 
%
%
%\bibitem[Curtis 
%\& Richer(2010)]{Curtis2010} Curtis, E.~I., et al.\ 2010, MNRAS, 402, 603 


%\bibitem[Donkov et al.(2011)]{Donkov2011} Donkov, S., et al.\ 2011, MNRAS, 418, 916 

\bibitem[\protect\citeauthoryear{Fabbiano et al.}{2019}]{2019ApJ...870...69F} Fabbiano G., et al., 2019, ApJ, 870, 69 

\bibitem[\protect\citeauthoryear{Federrath, Klessen, \& Schmidt}{2008}]{2008ApJ...688L..79F} Federrath C. et al., 2008, ApJ, 688, L79 

\bibitem[\protect\citeauthoryear{Federrath}{2015}]{2015MNRAS.450.4035F} Federrath C., 2015, MNRAS, 450, 4035 

\bibitem[\protect\citeauthoryear{Federrath et al.}{2016}]{2016ApJ...832..143F} Federrath C., et al., 2016, ApJ, 832, 143 

\bibitem[\protect\citeauthoryear{Federrath}{2018}]{2018PhT....71f..38F} Federrath C., 2018, PhT, 71, 38 


\bibitem[\protect\citeauthoryear{Ferri{\`e}re}{2001}]{2001RvMP...73.1031F} Ferri{\`e}re K.~M., 2001, RvMP, 73, 1031 

\bibitem[Genzel et al.(2010)]{2010RvMP...82.3121G} Genzel, R., et al.\ 2010, RvMP, 82, 3121 


\bibitem[Goodman et al.(1998)]{Goodman1998} Goodman, A.~A., et al.\ 1998, ApJ, 504, 223 

\bibitem[\protect\citeauthoryear{Guainazzi et al.}{1998}]{1998MNRAS.301L...1G} Guainazzi M., et al., 1998, MNRAS, 301, L1 

\bibitem[Heyer 
\& Dame(2015)]{Heyer2015} Heyer, M., et al.\ 2015, ARA\&A, 53, 583 

%\bibitem[K{\"o}nyves et 
%al.(2015)]{2015A&A...584A..91K} K{\"o}nyves, V., et al.\ 2015, A\&A, 584, A91 


\bibitem[\protect\citeauthoryear{Kauffmann et al.}{2017}]{2017A&A...603A..89K} Kauffmann J. et al., 2017, A\&A, 603, A89 

\bibitem[\protect\citeauthoryear{Khabibullin et al.}{2019}]{2019K} Khabibullin I. et al., 2019 (in prep.) 

\bibitem[\protect\citeauthoryear{Khullar et al.}{2019}]{2019arXiv190200934K} Khullar S., Krumholz M.~R., Federrath C., Cunningham A.~J., 2019, arXiv, arXiv:1902.00934 


\bibitem[Koyama et al.(1996)]{1996PASJ...48..249K} Koyama, K., et al.\ 
1996, PASJ, 48, 249 

\bibitem[\protect\citeauthoryear{Krivonos et al.}{2017}]{2017MNRAS.468.2822K} Krivonos R., et al., 2017, MNRAS, 468, 2822 


\bibitem[\protect\citeauthoryear{Kruijssen, Dale, \& Longmore}{2015}]{2015MNRAS.447.1059K} Kruijssen J.~M.~D., Dale J.~E., Longmore S.~N., 2015, MNRAS, 447, 1059 



\bibitem[McKee 
\& Ostriker(2007)]{McKee2007} McKee, C.~F. \& Ostriker \ 2007, ARA\&A, 45, 565 


%\bibitem[Mills 
%\& Battersby(2017)]{Mills2017} Mills, E.~A.~C., et al.\ 2017, ApJ, 835, 76 

\bibitem[\protect\citeauthoryear{Mac Low \& Klessen}{2004}]{2004RvMP...76..125M} Mac Low M.-M., Klessen R.~S., 2004, RvMP, 76, 125 


\bibitem[Molaro et al.(2016)]{2016A&A...589A..88M} Molaro, M., et al.\ 2016, A\&A, 589, A88 

\bibitem[\protect\citeauthoryear{Mori et al.}{2015}]{2015ApJ...814...94M} Mori K., et al., 2015, ApJ, 814, 94 

\bibitem[Muno et al.(2007)]{2007ApJ...656L..69M} Muno, M.~P., et al.\ 2007, 
ApJ, 656, L69 

\bibitem[\protect\citeauthoryear{Onus, Krumholz, \& Federrath}{2018}]{2018MNRAS.479.1702O} Onus A., Krumholz M.~R., Federrath C., 2018, MNRAS, 479, 1702 


\bibitem[\protect\citeauthoryear{Padoan et al.}{2014}]{2014prpl.conf...77P} Padoan P., Federrath C., Chabrier G., Evans N.~J., II, Johnstone D., J{\o}rgensen J.~K., McKee C.~F., Nordlund {\AA}., 2014, Protostars and Planets VI, Henrik Beuther, Ralf S. Klessen, Cornelis P. Dullemond, and Thomas Henning (eds.), University of Arizona Press, Tucson, 77 


\bibitem[Ponti et al.(2010)]{2010ApJ...714..732P} Ponti, G., et al.\ 2010, 
ApJ, 714, 732 


%\bibitem[Rathborne, Jackson, 
%\& Simon(2006)]{2006ApJ...641..389R} Rathborne, J.~M., et al.\ 2006, ApJ, 641, 389 


\bibitem[\protect\citeauthoryear{Rathborne et al.}{2015}]{2015ApJ...802..125R} Rathborne J.~M., et al., 2015, ApJ, 802, 125

\bibitem[\protect\citeauthoryear{Revnivtsev et al.}{2004}]{2004A&A...425L..49R} Revnivtsev M.~G., et al., 2004, A\&A, 425, L49 


%\bibitem[Roman-Duval et al.(2010)]{2010ApJ...723..492R} Roman-Duval, J., et 
%al.\ 2010, ApJ, 723, 492 


%\bibitem[Schmiedeke et 
%al.(2016)]{2016A&A...588A.143S} Schmiedeke, A., et al.\ 2016, A\&A, 588, A143 


\bibitem[Sunyaev et al.(1993)]{1993ApJ...407..606S} Sunyaev, R.~A., et al.\ 1993, ApJ, 407, 606 

\bibitem[Sunyaev \& Churazov (1996)]{1996AstL...22..648S} Sunyaev R.~A., Churazov E.~M., 1996, AstL, 22, 648 


\bibitem[\protect\citeauthoryear{Sunyaev \& Churazov}{1998}]{1998MNRAS.297.1279S} Sunyaev R., Churazov E., 1998, MNRAS, 297, 1279 

\bibitem[\protect\citeauthoryear{Terrier et al.}{2010}]{2010ApJ...719..143T} Terrier R., et al., 2010, ApJ, 719, 143 

\bibitem[\protect\citeauthoryear{Vazquez-Semadeni}{1994}]{1994ApJ...423..681V} Vazquez-Semadeni E., 1994, ApJ, 423, 681 


\bibitem[\protect\citeauthoryear{Uehara et al.}{2017}]{2017IAUS..322..162U} Uehara K. et al., 2017, IAUS, 322, 162


\bibitem[\protect\citeauthoryear{Walch et al.}{2015}]{2015MNRAS.454..238W} Walch S., et al., 2015, MNRAS, 454, 238 

\bibitem[Ward-Thompson et al.(2007)]{Ward2007} Ward-Thompson, 
D., et al.\ 2007, Protostars and Planets V, B. Reipurth, D. Jewitt, and K. Keil (eds.), University of Arizona Press, Tucson, 951 pp., 2007., p.33-46


%\bibitem[Ward-Thompson et al.(2016)]{Ward2016} Ward-Thompson, 
%D., et al.\ 2016, MNRAS, 463, 1008 


\bibitem[Williams et al.(2000)]{Williams2000} Williams, J.~P., et al.\ 2000, Protostars and Planets IV (Book - Tucson: University of Arizona Press; eds Mannings, V., Boss, A.P., Russell, S. S.), p. 97, 97 

\end{thebibliography}
\end{document}